# The Selective Transfer of Patterned Graphene

Xu-Dong Chen[1], Zhi-Bo Liu[1,*], Wen-Shuai Jiang[1], Xiao-Qing Yan[1,2], Fei Xing[1], Peng Wang[1], Yongsheng Chen[2], Jian-Guo Tian[1,*]

[1] The Key Laboratory of Weak Light Nonlinear Photonics, Ministry of Education, Teda Applied Physics School and School of Physics, Nankai University, Tianjin 300457, China

[2] The Key Laboratory of Functional Polymer Materials and Center for Nanoscale Science & Technology, Institute of Polymer Chemistry, College of Chemistry, Nankai University, Tianjin 300071, China

*email: rainingstar@nankai.edu.cn, jjtian@nankai.edu.cn

Abstract: We demonstrate a selective microcleaving graphene (MG) transfer technique for the transfer of graphene patterns and graphene devices onto chosen targets using a bilayer-polymer structure and femtosecond laser microfabrication. In the bilayer-polymer structure, the first layer is used to separate the target graphene from the other flakes, and the second layer transfers the patterned graphene to the chosen targets. This selective transfer technique, which exactly transfers the patterned graphene onto a chosen target, leaving the other flakes on the original substrate, provides an efficient route for the fabrication of MG for microdevices and flexible electronics and the optimization of graphene's performance. This method will facilitate the preparation of van der Waals heterostructures and enable the optimization of the performance of graphene hybrid devices.

## Introduction

Graphene is an emerging class of two-dimensional (2D) material with unique electrical properties[1-3] and a wide range of potential practical applications.[4,5] In addition, graphene hybrid structures combined with other 2D materials, metal microstructures, silicon photonic crystal cavities, and waveguides have more extensive applications in van der Waals heterostructures,[6-8] hybrid graphene plasmonics,[5,9] hybrid optoelectronic devices,[10,11] and optical modulators.[12,13] Based on well-developed transfer methods,[14-22] graphene grown by chemical vapor deposition (CVD) is currently used in most of the graphene hybrid applications.[11-13,23] Although mechanical exfoliation of highly oriented pyrolytic graphite provides the highest-quality graphene,[1,2] the transfer of the desired microcleaving graphene (MG) to the structure at a specific position is a critical challenge, that limits the combination of MG with other structures.

Currently, only a few methods have been reported for the transfer of MG.[6] In the commonly used MG transfer method, all the graphene/graphite flakes obtained by microcleaving are transferred to new substrates with a layer of poly(methyl methacrylate) (PMMA).[24] With this method, the unwanted graphene/graphite flakes are transferred to new substrates, along with the useful graphene. These unwanted graphene/graphite flakes and the irregular shape of the desired graphene may result in a number of issues: (1) the irregularly shaped graphene may deleteriously affect the performance of graphene hybrid device; (2) the flakes will inevitably contaminate the new substrates and may damage other samples and/or devices on the substrates; and

(3) the flakes may hinder the integration of multiple graphene devices. Therefore, the development of a selective transfer technique that enables the exact transfer of the useful patterned graphene to a chosen target while the other flakes are retained on the original substrate is important. In particular, the preparation of van der Waals heterostructures by the stacking of different 2D crystals on top of each other[6-8] requires that the crystals be transferred precisely while their beneficial qualities are maintained. If such a MG transfer technique is developed, we could replace CVD-grown graphene with MG in graphene-silicon photonic crystal cavity structures[11,23] and graphene-based optical modulators[12,13] and enhance the performance of these graphene hybrid optoelectronic devices.

In addition, the observation of a clear color contrast requires a $SiO_2$/Si substrate with a specific $SiO_2$ thickness,[1,25] which limits the fabrication of graphene devices on other types of substrates. The fabrication of samples and devices on other substrates is relevant for fundamental research and for the optimization of graphene's performance. In addition, the newly developed method should integrate graphene patterns or devices with a wider variety of substrate materials. Herein, we report a new technique for the selective transfer of MG patterns and devices onto chosen targets using a bilayer-polymer structure and femtosecond laser microfabrication. This selective transfer technique, which exactly transfers the patterned graphene onto a chosen target, leaving the other flakes on the original substrate, provides an efficient route for the fabrication of MG-based microdevices. A clean surface without any other flakes, with any desired shape for each staking layer, and with precise transfer achieved via this technique will facilitate the development of van der Waals heterostructures in electronics and devices.[6,8]

**Results**

Figure 1 illustrates the selective transfer process (for details, see Methods). The graphene patterns and devices were fabricated on $SiO_2$/Si wafers, and a layer of positive photoresist was subsequently coated onto the wafers. After the exposure and development, the patterned graphene was separated from the other graphene/graphite flakes. The femtosecond laser was used to fabricate graphene patterns and expose the photoresist above the patterned graphene. Because of their ultra-short pulse and high energy, focused femtosecond lasers are widely used in microfabrication to fabricate patterns with high resolution. Direct-laser-writing optical lithography[26,27] makes the exposure process more flexible, and the two-photon exposure can improve the resolution of the exposed patterns. The entire sample was subsequently exposed to a xenon lamp, and then a layer of PMMA was coated onto the sample as the transfer layer. As a support layer, PMMA is widely used in the transfer of CVD-grown graphene and carbon nanotubes.[14,20,28] In this work, PMMA was chosen as the transfer layer because of its relatively low viscosity and good wetting capability, which makes the PMMA solution conform well to nanoscale topography on $SiO_2$/Si substrate.[29] The baked PMMA film makes full contact with the patterned graphene, enhancing the van der Waals interactions between the patterned graphene and PMMA film. The enhanced van der Waals interaction is usually larger than that between graphene and $SiO_2$/Si substrate. Therefore, the patterned graphene was separated from the $SiO_2$/Si substrate and attached to the PMMA layer when the photoresist was dissolved, leaving the unwanted flakes on the SiO$_2$/Si substrate. Finally, the patterned graphene was precisely transferred to a chosen target with a micromanipulator mounted on an optical microscope and then the PMMA was dissolved in acetone for several times. After transferred, the sample was annealed in flowing H$_2$/Ar gas at 350 °C for 3 h to remove the residual photoresist and PMMA on graphene.[6,30] Figure S1 shows the AFM images of three few-layer graphene flakes transferred selectively onto flat SiO$_2$/Si wafers before and after annealing. After the thermal annealing process, most of the residues were removed, and graphene was in better contact with the substrates.[31]

In order to demonstrate the detachment yield of the graphene from SiO$_2$/Si substrate, 100 graphene sheets on 10 samples were transferred selectively to other substrates. 95 graphene sheets were detached from the original substrates and 93 sheets were transferred to other substrates successfully. Using this transfer method, high yields can up to 93%.

The selective transfer of the patterned graphene to the chosen target can be achieved with this method. A monolayer graphene was patterned and transferred to a chosen microcavity, as shown in Figure 2. Five cubic microcavities were fabricated on the SiO$_2$/Si substrate, and the patterned graphene was precisely transferred onto the center cavity (Figure 2b), leaving the other flakes on the original substrate for reuse (Figure 2c). The morphological images of the patterned graphene transferred by the selective transfer technique were analyzed using scanning electron microscopy (SEM) and atomic force microscopy (AFM). The SEM image (Figure 2d) shows that the transferred graphene film is clean and has a relatively uniform morphology without any apparent defects. The AFM image (Figure 2e), which has few apparent residual, further reveals the high quality of the transferred graphene. To demonstrate the integrity of the graphene above the microcavity, a 2D- to G-band intensity ratio map is shown in Figure 2f.[32,33] The Raman spectra of the graphene above the microcavity have a higher 2D/G ratio and a much lower intensity compared to those of the surrounding graphene (Supplementary Figure S2). The Raman image shows that the graphene is present at every point above the microcavity, thereby indicating good integrity. The MG-microcavity structure can be used in selective molecular sieving,[34] and this selective transfer technique provides an approach to the fabrication of suspended graphene membranes[34,35] that do not contain undesirable flakes.

The fabrication of graphene samples and devices on different types of substrates[4,36] is relevant for fundamental research and the optimization of graphene's performance. The identification of graphene with a clear color contrast requires a $SiO_2$/Si wafer with a specific $SiO_2$ thickness,[1,25] which limits the fabrication of graphene devices on other types of substrates, such as quartz and polyethylene terephthalate (PET). The transfer of as-fabricated micro/nano-patterns would facilitate the development of graphene-based, transparent, and flexible microelectronics and devices.[36,37] To demonstrate the pattern transfer, we fabricated graphene ribbons using a femtosecond laser and transferred them to other substrates. Figure 3 shows the SEM images of the graphene micro/nanoribbons on a $SiO_2$/Si substrate after the selective transfer process. Graphene ribbons with various widths (100 nm ~ 2 μm) were well printed to the new substrate without distortion or fracture. The minimum resolution (~ 100 nm) was limited by the objective lens and the motorized stage used in the experiment. We performed further verification by transferring different graphene patterns with different thicknesses onto various substrates (Supplementary Figure S3). The successful transfer suggests that the newly developed method can integrate graphene patterns with a wide variety of substrate materials.

The fabrication of electronics and devices is complex,[1] and graphene may be contaminated or even damaged during the fabrication process. An alternative approach is to apply all the fabrication steps, such as lithography, deposition, and lift-off, prior to the selective transfer of the graphene films with designed shapes fabricated on $SiO_2$/Si to the target position. Therefore, a graphene/electrodes structure can effectively avoid contamination and damage. A patterned monolayer graphene film was transferred selectively to as-fabricated Au electrodes on a $SiO_2$/Si wafer to fabricate a graphene field-effect transistor (FET) (Figure 4a). The SEM image (Figure 4d) shows that the graphene was well transferred and precisely located on the Au electrodes. The Raman spectra of the monolayer graphene on the $SiO_2$/Si (Figure 4e) and Au electrodes (Supplementary Figure S4) measured after the transfer process suggest that high-quality graphene was produced.[21,32] For comparison, we fabricated another graphene FET with an electrodes/graphene structure by depositing Au electrodes onto a graphene film. In this case, the graphene was found to be seriously contaminated (Supplementary Figure S5). Prior to each measurement, the samples were annealed at 200 ℃ for 3 h in vacuum to remove water and air doped onto the graphene. Figure 4f shows the room-temperature transport characteristics of the two FETs after they were annealed. The FET fabricated with the selective transfer method generated a narrower curve compared to that of the other FET. In addition, the charge neutrality points of the FETs fabricated by the selective transfer and conventional methods were 7 V and 10 V, respectively. In comparison to the conventional FET, the FET fabricated via selective transfer exhibited better performance due to less contamination. To investigate the devices transfer process, the FET fabricated via the selective transfer technique on the $SiO_2$/Si wafer was transferred to quartz using PMMA.[28] The optical images of the device on $SiO_2$/Si, PMMA, and quartz (Figures 4a, b, and c) show that the device was transferred with high fidelity. Figure 4g shows the room-temperature current as a function of the voltage of the graphene device on $SiO_2$/Si, PMMA, and quartz. The increase in the resistance was most likely caused by mechanical damage of the Au electrodes during the transfer process. With the newly developed method, more complex microdevices can be fabricated, such as an electrodes/graphene/hexagonal-boron-nitride (hBN) /electrodes/graphene structure.

Most microdevices are fabricated on rigid, flat, and smooth substrates,[1,2] such as $SiO_2$/Si. However, interest in nonconventional substrates, such as soft plastics, has been increasing because of these substrates' potential applications in transparent, flexible, or foldable electronics.[36,37] After confirming that the selective transfer technique could be used to fabricate electronics and devices on a $SiO_2$/Si wafer, we selected a PET film as the substrate for the fabrication of electronics via this method.

We fabricated ITO electrodes on the PET film using photolithography (see Methods) and then transferred a patterned monolayer graphene film to connect the two ITO films (Figures 5a and b). The Raman spectrum (Figure 5c) of the graphene measured on the PET film demonstrates the high quality of the graphene used in the experiment.[32] The inset of Figure 5b shows the SEM image of the transferred graphene. The patterned graphene is attached to the ITO films on PET, connecting the two electrodes. For comparison, we fabricated three other devices in which the ITO electrodes were connected by CVD-grown monolayer graphene, multilayer MG, and ITO. The resistance of the four devices with different bending radii and flat-bent cycles are shown in Figures 5d and 5e, respectively. The current as a function of the voltage for the four devices measured during the bending test is shown in Supplementary Figure S6. The increase in the resistance was caused primarily by the ITO films while the devices were bent multiple times to achieve different radii of curvature. Whereas the resistance variation of the ITO films was eliminated, the resistance of the MG-based devices was nearly invariant (see insets of Figures 5d and e). The results of the bending test suggest that MG films were successfully attached to the PET/ITO films, and that the MG films exhibited better mechanical stability than CVD-grown graphene.[14]

## Discussion

The developed selective transfer technique provides an efficient route for the fabrication of MG-based microdevices and flexible electronics. In addition, the selective transfer technique can be readily incorporated into the process used to fabricate van der Waals heterostructures. In general, van der Waals heterostructures are fabricated by stacking different 2D crystals on top of each other with a "mechanical transfer process".[6-8] However, with this method, the unwanted graphene/graphite flakes are transferred together with the useful graphene. The flakes inevitably contaminate the new substrates and damage other samples on the substrates. However, with our newly developed method, the selective transfer of 2D crystals with regular shapes can be achieved, and different 2D crystals can be positioned over each other with micrometer accuracy in a desired sequence. The transfer process, which is free of contaminants, enables the integration of multiple graphene devices. In addition, the regular shapes of the 2D crystals can improve the performance of van der Waals heterostructures. A clean surface without any other flakes, the ability to fabricate any desired shape for each staking layer, and the precise transfer obtained via this selective transfer technique will facilitate the development of van der Waals heterostructures in electronics and devices.[6,8]

In summary, we demonstrated a unique selective transfer technique using a photoresist/PMMA structure and femtosecond laser microfabrication to achieve the selective transfer of MG patterns onto chosen targets. With this method, as-fabricated graphene micro/nanoribbons were transferred onto different substrates, and graphene patterns were transferred onto different electrodes fabricated on various substrates. This method provides an efficient route for the fabrication of MG-based microdevices and flexible electronics and for the optimization of the performance of graphene device. We believe that this technique can also provide a feasible approach to the fabrication of van der Waals heterostructures with a clean surface, with any desired shape for each staking layer, and with precise positioning.

# Methods

**Materials.** Silicon wafers coated in a 285-nm thermal oxide were purchased from Graphene-Supermarket (graphene-supermarket.com). The positive photoresist (AR-P 3510T) and developer (AR 300-26) were obtained from Allresist. The PMMA (average $M_w$~ 996,000 by GPC, crystalline) was purchased from Sigma-Aldrich. The PET/ITO film (thickness: 0.125 mm, resistivity: 45 ± 5 ohm/sq, transmittance > 84 %) was purchased from Zhuhai Kaivo Electronic Components.

**Selective transfer process.** Graphene patterns were fabricated using a focused femtosecond laser (~ 100 fs, 800 nm, 80 MHz) after the deposition and recognition of the MG on the $SiO_2$/Si wafers. A layer of positive photoresist was spin-coated (3000 rpm, 30 s) onto the substrate, and the coated substrate was prebaked at 100 ℃ for 2 min. An 800-nm femtosecond laser was used to expose the patterned graphene region via a two-photo exposure. The exposed photoresist was dissolved after development to reveal the patterned graphene. The entire sample was subsequently exposed using a 150 W xenon lamp for 4 min, and a layer of PMMA was spin-coated (2000 rpm, 30 s) onto the sample. The patterned graphene was attached to the PMMA layer, and the other graphene/graphite flakes were separated by the photoresist layer. The sample was rinsed in the developer to dissolve the photoresist layer. The PMMA membrane could be peeled from the $SiO_2$/Si substrate when the photoresist was partially dissolved, and the unwanted flakes remained on the substrate. After being rinsed with deionized water, the PMMA membrane was lifted up by the new substrate. The position of the patterned graphene on the PMMA membrane was determined using a microscopy, and the patterned graphene was precisely aligned to the target position with a micromanipulator. The deionized water between the PMMA membrane and substrate protected the patterned graphene and substrate during the location process. After the deionized water was removed, the patterned graphene was attached to the target position. The sample was baked at 150 °C for 15 min to evaporate the residual water and improve the contact between the patterned graphene and the substrate. After the graphene was transferred, the PMMA was dissolved in acetone, and the sample was subsequently annealed in flowing $H_2$/Ar gas at 350 °C for 3 h to remove the resist residues. Using this technique, we could position the graphene on a chosen target within a few micrometers.

**Microcavity fabrication.** Five squares with a side length of 15 μm were defined by femtosecond laser direct writing lithography on $SiO_2$/Si. Then, a buffered oxide etch (BOE) solution was used to etch the squares into cubic microcavities with a depth of ~ 100 nm.

**Device fabrication.** The PET/ITO/graphene devices were fabricated using a combination of photolithography and the selective transfer process. The ITO film on PET (15 mm × 20 mm) was divided into two equal portions by photolithography and etching. The width of the gap between the two ITO films was 3 μm. The rectangular graphene was subsequently transferred to the middle of the gap to connect the two

ITO films. During the bending tests, indium-eutectic was used as a contact electrode to reduce contact resistance.

**Characterization.** The surface morphologies of graphene patterns and devices were investigated using optical microscopy (Nikon ECLIPSE Ti-U), AFM (Nanoscope Dimension$^{TM}$ 3100) and SEM (LEO1530VP). Raman spectra were obtained using a RENISHAW RM2000 Raman system equipped with a 514 nm laser source and 50× objective lens. To obtain the Raman images, the samples were moved with a step size of 2 μm, and a Raman spectrum was recorded at each point. The measured region was 22 × 22 μm$^2$. The transfer characteristic curve and current as a function of the voltage for the devices were measured using an Agilent 34401A Digit Multimeter and a Keithley 2400 SourceMeter.

# References

1    Novoselov, K. S. *et al.* Electric field effect in atomically thin carbon films. *Science* **306**, 666-669, (2004).

2    Novoselov, K. S. *et al.* Two-dimensional gas of massless Dirac fermions in graphene. *Nature* **438**, 197-200, (2005).

3    Avouris, P. Graphene: Electronic and Photonic Properties and Devices. *Nano Lett.* **10**, 4285-4294, (2010).

4    Novoselov, K. S. *et al.* A roadmap for graphene. *Nature* **490**, 192-200, (2012).

5    Bao, Q. & Loh, K. P. Graphene Photonics, Plasmonics, and Broadband Optoelectronic Devices. *ACS Nano* **6**, 3677-3694, (2012).

6    Dean, C. R. *et al.* Boron nitride substrates for high-quality graphene electronics. *Nat. Nanotech.* **5**, 722-726, (2010).

7    Geim, A. K. & Grigorieva, I. V. Van der Waals heterostructures. *Nature* **499**,

419-425, (2013).

8   Dean, C. *et al.* Graphene based heterostructures. *Solid State Commun.* **152**, 1275-1282, (2012).

9   Grigorenko, A. N. *et al.* Graphene plasmonics. *Nat. Photon.* **6**, 749-758, (2012).

10  Bonaccorso, F. *et al.* Graphene photonics and optoelectronics. *Nat. Photon.* **4**, 611-622, (2010).

11  Gu, T. *et al.* Regenerative oscillation and four-wave mixing in graphene optoelectronics. *Nat. Photon.* **6**, 554-559, (2012).

12  Liu, M. *et al.* A graphene-based broadband optical modulator. *Nature* **474**, 64-67, (2011).

13  Liu, M. *et al.* Double-Layer Graphene Optical Modulator. *Nano Lett.* **12**, 1482-1485, (2012).

14  Li, X. *et al.* Transfer of Large-Area Graphene Films for High-Performance Transparent Conductive Electrodes. *Nano Lett.* **9**, 4359-4363, (2009).

15  Kim, K. S. *et al.* Large-scale pattern growth of graphene films for stretchable transparent electrodes. *Nature* **457**, 706-710, (2009).

16  Lee, Y. *et al.* Wafer-Scale Synthesis and Transfer of Graphene Films. *Nano Lett.* **10**, 490-493, (2010).

17  Bae, S. *et al.* Roll-to-roll production of 30-inch graphene films for transparent electrodes. *Nat. Nanotech.* **5**, 574-578, (2010).

18  Kulkarni, A. *et al.* A Novel Method for Large Area Graphene Transfer on the Polymer Optical Fiber. *J. Nanosci. Nanotechnol.* **12**, 3918-3921, (2012).

19  Kang, J. *et al.* Efficient Transfer of Large-Area Graphene Films onto Rigid Substrates by Hot Pressing. *ACS Nano* **6**, 5360-5365, (2012).

20  Kang, J. *et al.* Graphene transfer: key for applications. *Nanoscale* **4**, 5527-5537, (2012).

21  Chen, X.-D. *et al.* High-quality and efficient transfer of large-area graphene films onto different substrates. *Carbon* **56**, 271-278, (2013).

22  Song, J. *et al.* A general method for transferring graphene onto soft surfaces.

*Nat. Nanotech.* **8**, 356-362, (2013).

23  Majumdar, A. *et al.* Electrical Control of Silicon Photonic Crystal Cavity by Graphene. *Nano Lett.* **13**, 515-518, (2013).

24  Reina, A. *et al.* Transferring and Identification of Single- and Few-Layer Graphene on Arbitrary Substrates. *J. Phys. Chem. C* **112**, 17741-17744, (2008).

25  Blake, P. *et al.* Making graphene visible. *Appl. Phys. Lett.* **91**, (2007).

26  Deubel, M. *et al.* Direct laser writing of three-dimensional photonic-crystal templates for telecommunications. *Nat Mater* **3**, 444-447, (2004).

27  Fischer, J. & Wegener, M. Ultrafast Polymerization Inhibition by Stimulated Emission Depletion for Three-dimensional Nanolithography. *Adv. Mater.* **24**, OP65-OP69, (2012).

28  Thanh, Q. N. *et al.* Transfer-Printing of As-Fabricated Carbon Nanotube Devices onto Various Substrates. *Adv. Mater.* **24**, 4499-4504, (2012).

29  Jiao, L. *et al.* Creation of Nanostructures with Poly(methyl methacrylate)-Mediated Nanotransfer Printing. *J. Am. Chem. Soc.* **130**, 12612-12613, (2008).

30  Hunt, B. *et al.* Massive Dirac Fermions and Hofstadter Butterfly in a van der Waals Heterostructure. *Science* **340**, 1427-1430, (2013).

31  Xu, W. *et al.* Graphene-Veiled Gold Substrate for Surface-Enhanced Raman Spectroscopy. *Adv. Mater.* **25**, 928-933, (2013).

32  Ferrari, A. C. *et al.* Raman Spectrum of Graphene and Graphene Layers. *Phys. Rev. Lett.* **97**, 187401, (2006).

33  Nolen, C. M. *et al.* High-Throughput Large-Area Automated Identification and Quality Control of Graphene and Few-Layer Graphene Films. *ACS Nano* **5**, 914-922, (2011).

34  Koenig, S. P. *et al.* Selective molecular sieving through porous graphene. *Nat. Nanotech.* **7**, 728-732, (2012).

35  Koenig, S. P. *et al.* Ultrastrong adhesion of graphene membranes. *Nat. Nanotech.* **6**, 543-546, (2011).

36  Kaltenbrunner, M. *et al.* An ultra-lightweight design for imperceptible plastic electronics. *Nature* **499**, 458-463, (2013).

37  Lee, S.-K. *et al.* All Graphene-Based Thin Film Transistors on Flexible Plastic Substrates. *Nano Lett.* **12**, 3472-3476, (2012).

**Acknowledgements**

The authors wish to thank the Chinese National Key Basic Research Special Fund (grant 2011CB922003) and the Natural Science Foundation of China (grant 11174159).

**Author contributions**

X.D.C. and Z.B.L. performed the experiments and analyzed the data. X.Q.Y, F.X., and Y.C. analyzed the data. P.W. and W.S.J conceived and designed the experiments, and analyzed the data. X.D.C., Z.B.L., and J.G.T. co-wrote the paper. X.D.C., Z.B.L., and J.G.T. contributed to methodology development. All the authors discussed the results and commented on the manuscript.

**Additional information**

Supplementary information is available in the online version of the paper. Reprints and permissions information is available online at www.nature.com/reprints. Correspondence and requests for materials should be addressed to Z.B.L.

**Competing financial interests**

The authors declare no competing financial interests.

**Figures:**

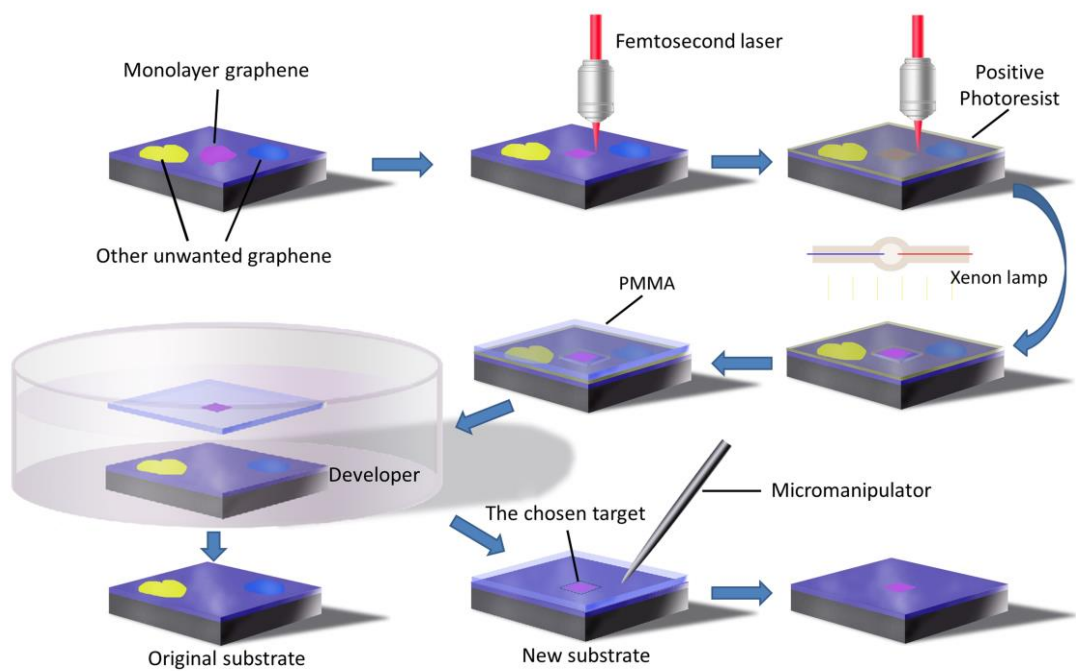

**Figure 1| Schematic illustrations of the selective transfer process.** A femtosecond laser at 800 nm was used to fabricate the graphene patterns and expose the photoresist. The micromanipulator mounted on an optical microscope was used to position the graphene patterns on the chosen targets.

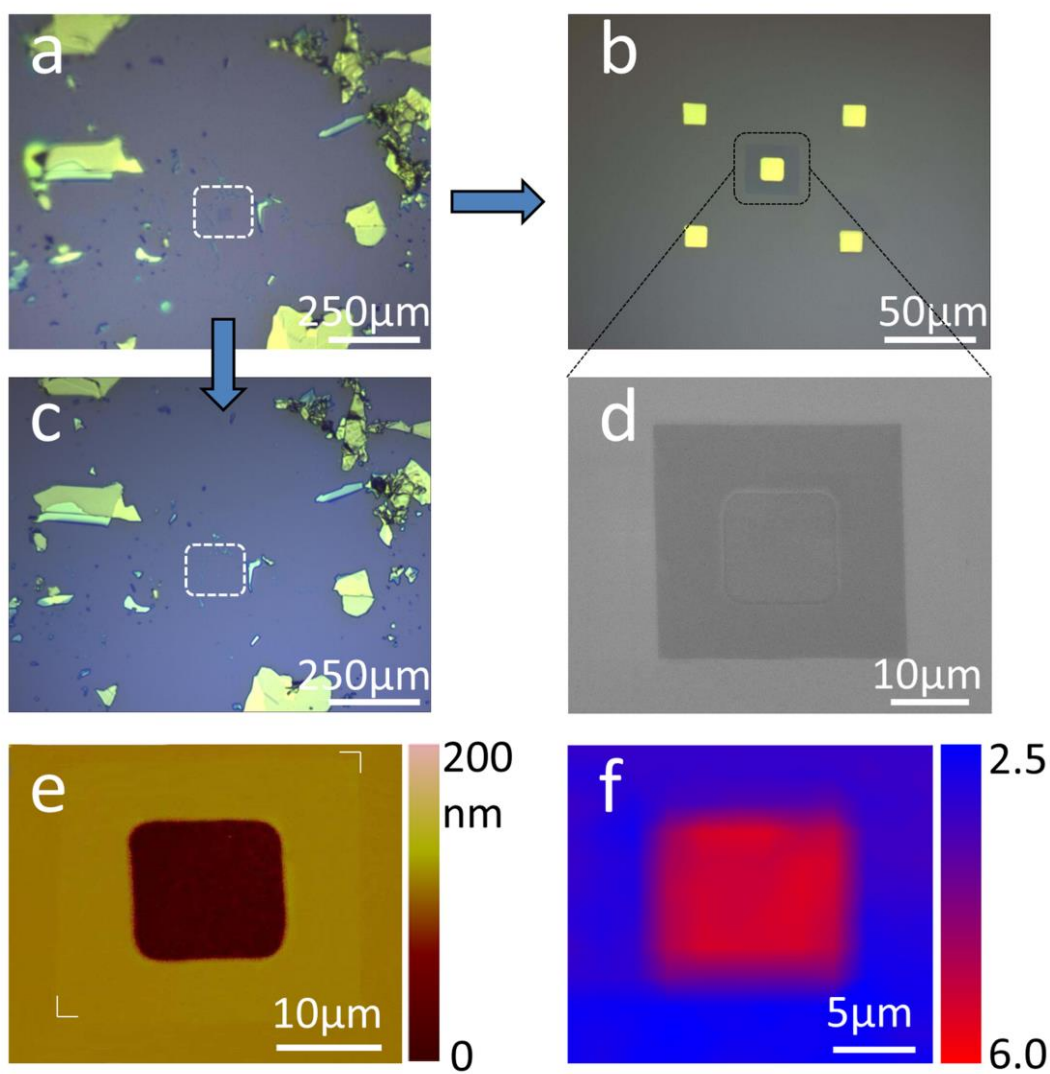

**Figure 2| Results of the selective transfer and precise location. a-c**, Optical images of (**a**) a monolayer MG patterned by femtosecond laser on a SiO$_2$/Si wafer, (**b**) the patterned graphene located on a chosen microcavity obtained by selective transfer, and (**c**) the graphene/graphite flakes remaining on the original substrate. **d-f**, The morphological images of the patterned graphene above the microcavity analyzed by (**d**) SEM, (**e**) AFM, and (**f**) Raman mapping with a 2D- to G-band intensity ratio.

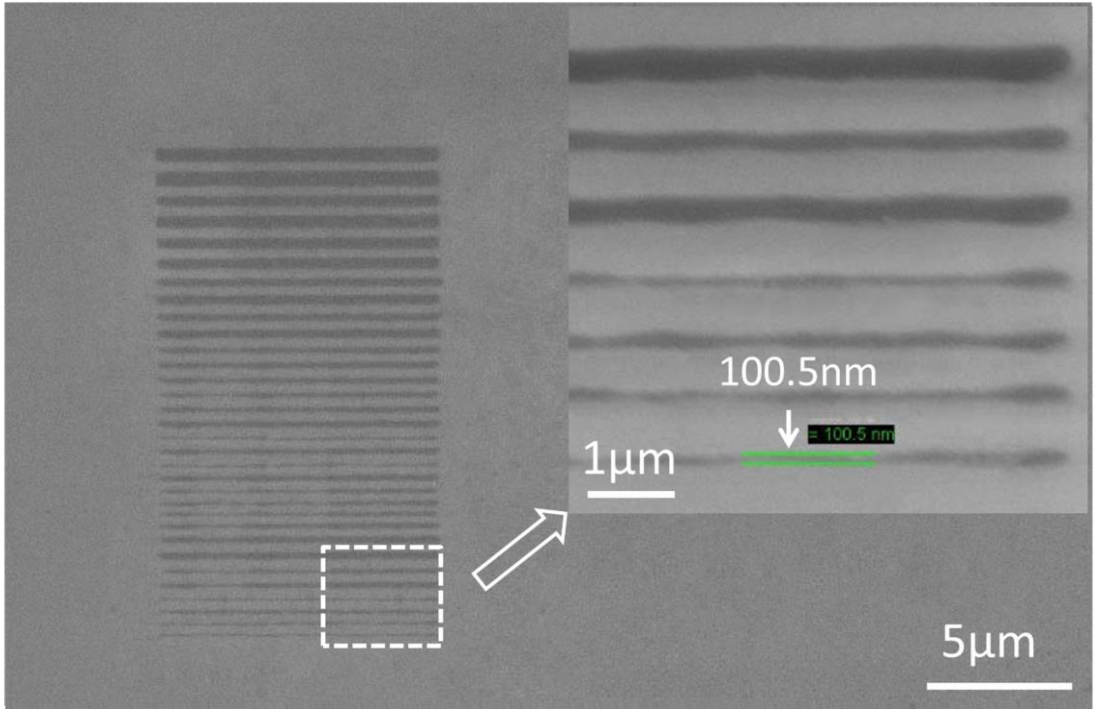

**Figure 3| SEM images of the graphene micro/nanoribbons.** Monolayer MG micro/nanoribbons were fabricated and selectively transferred with high fidelity.

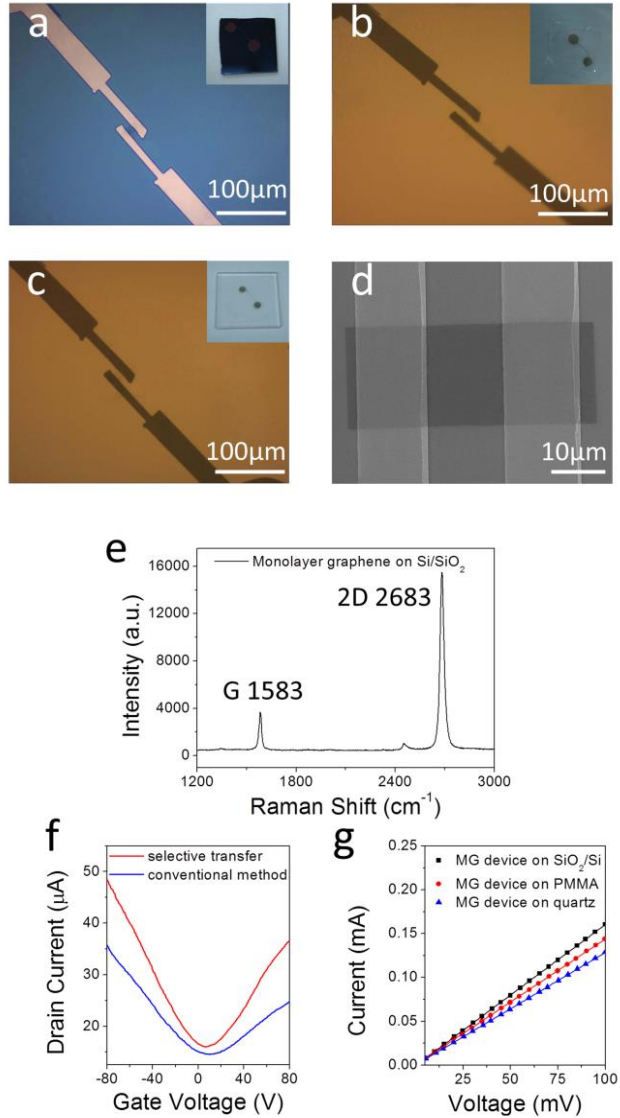

**Figure 4| The fabrication and transfer of the graphene FET with graphene/electrodes structure. a-c**, Optical images of the graphene device on (**a**) Si/SiO$_2$, (**b**) PMMA, and (**c**) quartz. **d**, SEM image of the patterned graphene transferred onto the Au electrodes to fabricate the device. **e**, The Raman spectrum of the transferred graphene on a Si/SiO$_2$ wafer. **f**, The transfer characteristic curves of the two graphene FETs fabricated via the selective transfer and conventional methods. **g**, The current as a function of the voltage for the graphene device fabricated by the selective transfer during the device transfer process.

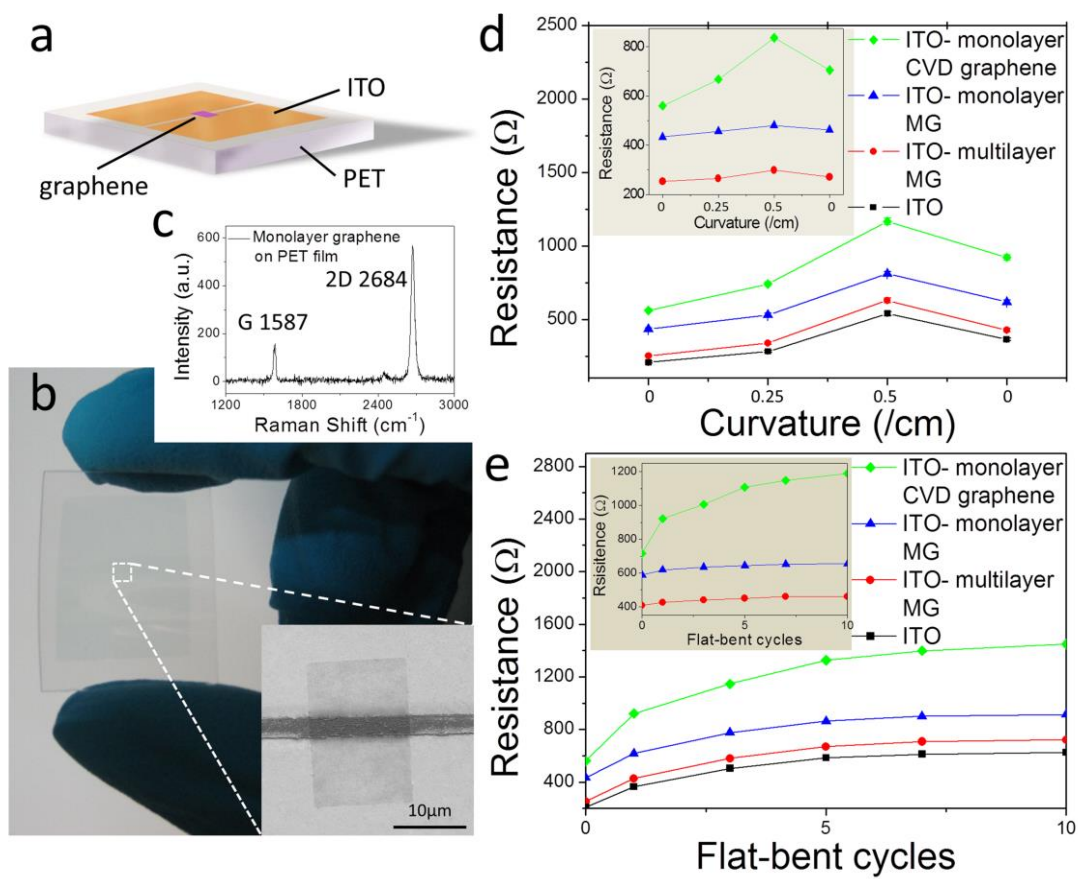

**Figure 5| Bending test of the PET/ITO/graphene flexible device fabricated by selective transfer.** (**a**) The configuration and (**b**) a photograph of the PET/ITO/graphene device. The inset in (**b**) is the SEM image of the transferred graphene connecting the two ITO films. **c**, The Raman spectrum of the transferred graphene on PET. **d**, The changes in the resistance of the four devices as a function of the bending radius. **e**, Resistance of the devices with different flat-bent cycles. The insets in (**d**) and (**e**) show the results with the resistance variation of the ITO films removed from (**d**) and (**e**), respectively.